\documentclass[manuscript]{aastex}

\newcommand{\HST}{{\it HST}}

\begin{document}          


\title{{\em Hubble Space Telescope\/} Snapshot Survey for Resolved Companions of
Galactic Cepheids\altaffilmark{1}} 

\author{
Nancy Remage Evans,\altaffilmark{2}
Howard E. Bond,\altaffilmark{3,4}
Gail H. Schaefer,\altaffilmark{5} 
Brian D. Mason,\altaffilmark{6} 
Evan Tingle,\altaffilmark{2}  
Margarita Karovska,\altaffilmark{2}  
and   
Ignazio Pillitteri\altaffilmark{2,7}  
}  

\altaffiltext{1}
{Based on observations with the NASA/ESA {\it Hubble Space Telescope\/} obtained
at the Space Telescope Science Institute, which is operated by the Association
of Universities for Research in Astronomy, Inc., under NASA contract
NAS5-26555.}

\altaffiltext{2}
{Smithsonian Astrophysical Observatory, MS 4, 60 Garden St., Cambridge, MA
02138, USA; nevans@cfa.harvard.edu}

\altaffiltext{3}
{Department of Astronomy \& Astrophysics, Pennsylvania State University,
University Park, PA 16802, USA; heb11@psu.edu}

\altaffiltext{4}
{Space Telescope Science Institute, 3700 San Martin Drive, Baltimore, MD
21218, USA}

\altaffiltext{5}
{The CHARA Array of Georgia State University, Mount Wilson Observatory,
Mount Wilson, CA 91023, USA}

\altaffiltext{6}
{US Naval Observatory, 3450 Massachusetts Ave., NW, Washington, DC 20392, USA}

\altaffiltext{7}
{INAF-Osservatorio di Palermo, Piazza del Parlamento 1,I-90134 Palermo, Italy}

\begin{abstract}

We have conducted an imaging survey with the {\it Hubble Space Telescope\/} Wide
Field Camera~3 (WFC3) of 70 Galactic Cepheids, typically within 1~kpc, with the aim of
finding  resolved physical companions. The WFC3 field typically covers the 0.1 pc area
where companions are expected. 
In this paper, we identify 39 Cepheids
having candidate companions, based on their positions in color--magnitude
diagrams, and having separations $\geq\!5\arcsec$ from the Cepheids. We use
follow-up observations of 14 of these candidates with {\it XMM-Newton\/}, and of
one of them with {\it ROSAT\/}, to separate X-ray-active young stars (probable
physical companions) from field stars (chance alignments).  Our preliminary
estimate, based on the optical and X-ray observations, is that only 3\% 
of the Cepheids in the sample have wide companions. Our survey easily detects
resolved main-sequence companions as faint as spectral type K\null. Thus the
fact that the two most probable companions (those of FF~Aql and RV~Sco)
are earlier than type K is not simply a function of the detection limit. We find
no physical companions having separations larger than 4,000~AU in the X-ray
survey. Two Cepheids are exceptions in that they do have young companions at
significantly larger separations ($\delta$~Cep and S~Nor), but both belong to a
cluster or a loose association, so our working model is that they are not
gravitationally bound binary members, but rather cluster/association  members. 
All of these properties provide constraints on both star formation  and
subsequent dynamical evolution.  The low frequency of true physical 
companions at separations $>\!5''$ is confirmed by examination of the
subset of the nearest Cepheids and also the
density of the fields.

%
%

%
%
%
%
%
%
%

\end{abstract}




\section{Introduction}


Particular interest is being paid at present to the role of binaries, for
instance in planet
formation. Binary/multiple systems  have important effects in stellar
evolution, especially of massive stars, which have a high fraction of binaries
(e.g., Sana et al.\ 2012). A substantial fraction of the massive stars in such systems
exchange mass or merge during post-main-sequence evolution. Binary systems are
also progenitors of X-ray sources in later stages when one component has
evolved into a compact object. 

An important topic currently under discussion is the question of the formation
of wide binaries. Binaries are considered wide when they have separations of
about 1,000 to 10,000~AU, corresponding to orbital periods of approximately
12,000 to 400,000 years.  As discussed by Kouwenhoven et al.\ (2010), the
typical size of a star-forming core is 10$^3$ AU, which is also roughly the
separation between protostars in young clusters. Forming multiple star systems
wider than this does not follow naturally from a collapse process. Several
scenarios have been proposed to get around this. Kouwenhoven et al.\ suggest
that distant companions can be acquired as a star cluster disperses.
Alternatively, Reipurth \& Mikkola (2012) propose that distant components can
result from triple systems that are formed as compact units, but then ``unfold''
because they are dynamically unstable, sending one component to a wider orbit
(or ejecting it from the system).  When discussing wide companions, it is of
course possible that at least one star is itself actually a closer binary, i.e.,
that the system is a hierarchical triple. In this discussion, we use ``binary''
as shorthand for ``binary or multiple.'' A distant component, of course, may
result from a mixture of these processes to augment the few wide systems formed
in the collapse process. 

The diversity of configurations in multiple systems provides 
insights into the evolution of the stellar population. Single stars can evolve
without outside influence. Stars in binary systems may interact when the primary
expands past the main sequence if their orbital separations are small enough.
At wider separations, the components of binaries follow the same evolutionary
paths as single stars. Triple or higher multiple systems may in addition have 
dynamical evolution within the system, which may result in an increased 
separation (typically for the smallest member) or even 
the ejection of a
member.  Thus, multiple systems provide a length (separation)
measurement.   In the case of triple systems, it  may be altered by the internal
dynamics. 
Finally, the widest systems are fragile and subject to destruction
by external passing stars. 
For all these reasons,  the assembly of binary characteristics, particularly as a
function of the mass of the primary, provides a tool to investigate both
formation conditions and subsequent interactions.


To probe these questions, we are making a series of studies of binary systems with a
range  of separation and mass ratio. Cepheid variables provide valuable
information about the binary/multiple characteristics of fairly massive stars
(typically $\sim\!6\, M_\odot$). One of our studies uses radial velocities from
the sharp lines of Cepheids to derive the properties of spectroscopic-binary
systems with periods between about 1 and 20~years (Evans et al.\ 2015).  To
provide data about the binary frequency among main-sequence B stars (destined to
become Cepheids), we discussed the {\it Chandra\/} observation of the cluster
Tr~16 (Evans et al.\ 2011).

In order to explore the widest orbital separations, we have also carried out a
survey for resolved companions of Cepheids, using the {\it Hubble Space
Telescope\/} (\HST) and its Wide Field Camera~3 (WFC3). Initial results from
this survey were included in Evans et al.\ (2013, hereafter ``Paper~I''), who
discuss the subset of the sample with relatively high-mass companions  derived
using {\it IUE\/} observations. Paper~I provides the distribution of separations
for systems with mass ratios $q = M_2/M_1 > 0.4$, in a sample equally sensitive
across the range of possible separations. The current paper (Paper~II) is the
second in this series.  

The present paper gives full details of our WFC3 imaging survey of 70 Cepheids,
which was introduced in Paper~I\null.  Because of the challenges of identifying
and measuring faint close companions of much brighter Cepheids, we have divided
the discussion according to the detection approach. In the current paper we
discuss candidate companions located $5''$ or more from the Cepheids.  For these
systems the light of the much brighter Cepheid does not materially affect
photometry of the companions. A subsequent paper (Paper~III) will report on
companions closer than $5''$, which are embedded in the wings of the image of
the bright Cepheid; the photometry thus requires a more sophisticated
point-spread function (PSF) subtraction.  Since the separation in AU depends
both on the apparent separation and the distance, we defer discussion of the
full sample to Paper~III, in which we will combine both separation regimes. 
Several features of the current study are of note. Information about resolved
companions has advanced greatly with instrumental developments such as adaptive
optics (AO) and interferometry, particularly at small separations.    The current
study provides a uniform survey of orbital separations from several hundred AU
up to $\sim$0.1~pc (including the systems discussed in Paper~I and those to be
discussed in Paper~III).  In this study, very low-mass stellar companions can be
detected, but we limit our discussion to stars hotter than M0 for reasons of
field contamination and X-ray followup.  
Important insight about the possible companions
identified in this paper is provided by {\it XMM-Newton\/} ({\it XMM}) X-ray
observations to identify low-mass stars young enough to be physical companions
of the Cepheids (Evans et al.\ 2016; Paper~IV).

This paper is organized as follows. First the sample of 70 Cepheids that were
observed with WFC3 is described. Within this sample we then identify candidate
companions with a separation greater than 5$\arcsec$ but still within the WFC3
$40''\times40''$ aperture, and having a magnitude and color appropriate for a
main-sequence star at the the distance and reddening of the Cepheid (Table~A1). 
For a subset of these candidates, we use {\it XMM-Newton\/} observations to
select those that have the X-ray emission expected for main-sequence stars with
the ages of Cepheids. Finally, we discuss the results in the context of the 
subset of the nearest Cepheids, and also field density.  





\section{{\em HST\/} Snapshot Survey}

\subsection{Observations}

We have carried out a snapshot imaging survey of nearby Galactic Cepheids, using
the UVIS channel of the WFC3 camera on \HST\null. Because of the high
sensitivity of WFC3 and the fact that our targets are very bright, we elected to
use intermediate-band filters instead of the broad-band ones used most often in
\HST\/ imaging. The two filters we chose were F621M and F845M\null. Magnitudes
in these filters can be transformed reasonably well into broad-band $V$ and $I$,
respectively. 

The program was carried out in \HST\/ Cycle~18 (program ID SNAP-12215, PI
N.R.E.), over the interval from 2010 September~20 to 2011 September~11. Our
input list of snapshot targets contained 71 Cepheids. Remarkably, all 71 targets
were observed (an unusually high yield for a snapshot program). However, due to
a spacecraft data formatter error, the data from one of the observations
(FM~Aql) were lost. Three additional observations suffered failures to acquire
one of the two guide stars, but the data are still useful. Our program therefore
successfully covered 70 Galactic Cepheids.

For each WFC3 observation, we centered the Cepheid in a $1024\times1024$-pixel
subarray. The WFC3 UVIS image scale is $0\farcs0396\,\rm pixel^{-1}$, giving a
field of view of $40''\times40''$. We obtained three dithered exposures in both
of the filters. Exposure times---generally just a few seconds---were chosen such
that the image of the Cepheid (at its average brightness) would be overexposed,
leading to  a few saturated pixels at the center of its image. This overexposure
strategy results in companion stars being detectable to faint levels to within a
few pixels of the primary star. The images are available from the Mikulski
Archive for Space Telescopes\footnote{\tt http://archive.stsci.edu}, as are the
exposure times in each filter and the dates of the observations.   

The 70 Cepheids that were imaged in the survey are listed in Table~1.  The input
target list was compiled using the Galactic Cepheid  database\footnote{\tt
http://www.astro.utoronto.ca/DDO/research\slash cepheids/table\_physical.html}
(Fernie et al.\ 1995). We chose primarily the nearest Cepheids, most of which are
within 1~kpc. To these, we added three more luminous long-period Cepheids at
larger distances (T~Mon, RS~Pup, and SV~Vul). The pulsation period,
intensity-weighted mean $V$ magnitude, and $E(B-V)$ listed in Table~1 were
generally taken directly from the database. The exceptions were a few Cepheids
with relatively bright companions, discussed in Paper~I\null. For these stars,
the values of $\langle V\rangle$ and $E(B-V)$ have been corrected for the light
of the companion. The data for Y~Car have been taken from Evans (1992), likewise
corrected to subtract the light of the companion.

To use the Leavitt (period-luminosity, or P-L) relation to derive a distance, we
must first identify those Cepheids that are not pulsating  in the fundamental
mode. These ``s-Cepheids'' were identified historically based on their
low-amplitude, nearly symmetric sinusoidal light curves (in contrast to the
``sawtooth'' light curves of fundamental pulsators). With the development of
Fourier decomposition, it was realized that various combinations of amplitude
and phase parameters for low-order modes could distinguish between fundamental
and overtone pulsators.  This interpretation was confirmed observationally with
Magellanic Cloud microlensing programs such as the MACHO project (Alcock et al.\
1995), which showed two P-L sequences.  While this framework is secure, there
are still a few exceptions or puzzles. We have used photometric results
(Antonello et al.\ 1990) to identify the following as overtone pulsators:
SU~Cas, IR~Cep, BP~Cir, AV~Cir, DT~Cyg, SZ~Tau, LR~TrA, and AH~Vel.  We have
added BG~Cru, MY Pup, and V950~Sco, based on results from radial-velocity curves
(Kienzle et al.\ 1999). In addition, V1334~Cyg is classified by Evans (2000) as
a first-overtone pulsator.  The pulsation mode of V440~Per has been
controversial, partly because it has a period of 7.57 days, long enough that
some of the diagnostics become confused. Recent velocity data, however, indicate
that it is an overtone pulsator (Baranowski et al.\ 2009), which we will
adopt.   FF~Aql, on the other hand, has many Fourier characteristics of overtone
pulsators.  However, the \HST\/  Fine Guidance Sensor (FGS) parallax (Benedict
et al.\ 2007) indicates that it pulsates in the fundamental mode, which we will
use.  

Two stars in the sample require further discussion. CO~Aur is a double-mode
Cepheid, which is excited in the first  and second overtones (Antonello et al.\
1986).  We have fundamentalized the first-overtone period (1.78~days)
 using the relation from Alcock et al.\ (1995), as
discussed above.  V473~Lyr is unique as a Population~I Cepheid with a large
variation in pulsation amplitude over a period of approximately 1210~days (Burki
\& Mayor 1980).   Its absolute magnitude and pulsation mode have been discussed
a number of times. While they are still open to question, the consensus is that
it is pulsating in the second overtone (e.g., Burki et al.\ 1986; Andrievsky et
al.\ 1998), which is in keeping with its very short period. We have used the
$P_2/P_1$ ratio from CO~Aur (0.8007) to derive the first-overtone period from
the observed period (1.49~days), and then fundamentalized that. As a point of
interest, in addition to CO~Aur,   the stars  Y~Car, TU~Cas, EW~Sct, and U~TrA 
are also double-mode pulsators.


Distances for the targets, listed in Table~1, were computed in the same way as
in Paper~I, based on the Leavitt relation derived from the \HST\/ FGS parallaxes
of Benedict et al.\ (2007). The periods in Table~1 for the overtone pulsators
have  been fundamentalized  before computing the distance.  As discussed in
Evans (1991), for broad-band colors, a Cepheid is less reddened than a hot star
by the same intervening material.  We compensate  for this by using  $R =
A_V/E(B-V) = 3.46$ to compute  $V_0$ as needed.  For these nearby Cepheids, the
effect of using this value of $R$ rather than the more standard 3.1 is generally
small.  






In order to provide a sense of the scope of the survey, Column~6 in Table~1
gives an estimate of the smallest separation at which companions will be
detected  when the full reductions are complete, including the PSF correction to
reveal companions within $5''$ of the Cepheid (Paper~III).  We
have used an  estimate of 0\farcs3 as the radius limit outside of which
companions can be detected.  Column~7 provides the separation in
arcseconds corresponding to 0.1~pc at the distance of the Cepheid.   Systems
with wider separations are thought to be disrupted by the Galactic tidal field,
or through encounters with passing stars.  Column~7 shows that 
a large part of this  0.1~pc zone is contained 
 within the $40''\times 40''$ WFC3 field of view for most targets.



\subsection{Data Reduction and Analysis}

\begin{figure}
 \includegraphics[width=\columnwidth]{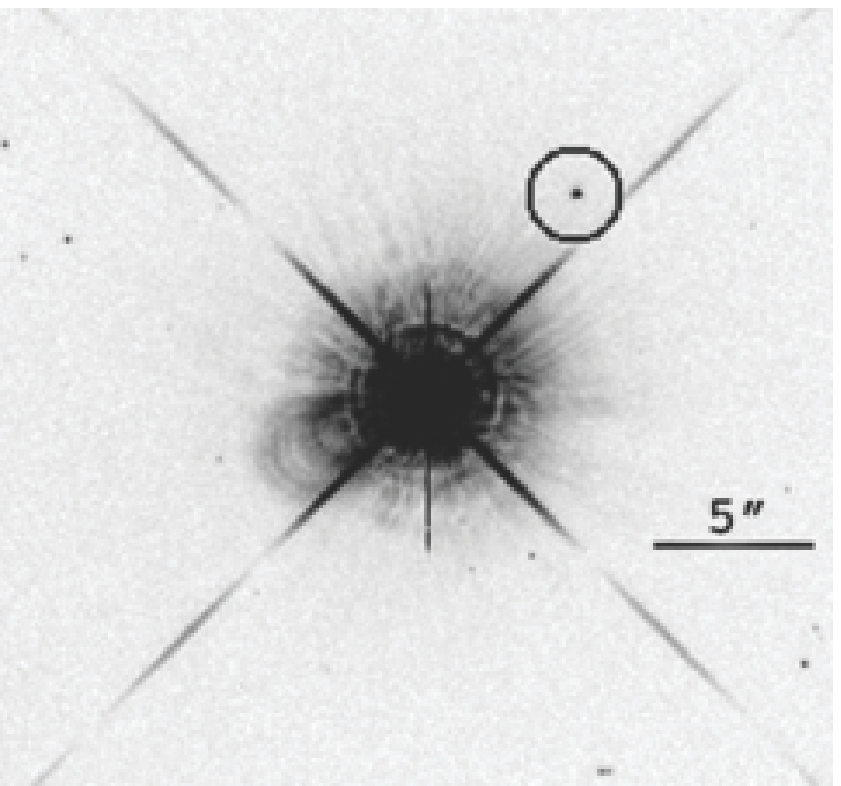}
\caption{R Cru image in the F845M filter.  The image has a log scale.  The 
possible companion is circled, and the spatial scale is indicated by the 
5$\arcsec$ bar.  \label{rcru.im}}
\end{figure}

As mentioned above, the analysis of the observations is divided into two parts. 
In this paper, possible companions more than 5$\arcsec$ from the Cepheid are
discussed. In such cases, standard aperture photometry can be used
straightforwardly.  Fig.~\ref{rcru.im} is a typical image, showing that light
from the  Cepheid contributes a complicated background to the image inside about
5$\arcsec$.

For the photometric analysis, we use the default drizzle-combined {\tt drz.fits}
images from the \HST\/ archive pipeline. These frames are created by combining
the individual dithered exposures, and are fully processed to bias-subtracted,
flat-fielded, and geometrically corrected images with cosmic rays removed.  In
the IRAF\footnote{IRAF is distributed by the National  Optical Astronomy
Observatories, which are operated by the Association of Universities for
Research in Astronomy, Inc., under cooperative agreement with the National
Science Foundation.}  environment, the \texttt{daofind}  routine was used to
locate all detected stars in the frames, and measure their image coordinates. 
When \texttt{daofind}  misses targets, a manual examination of the field using
\texttt{imexamine} provides the remaining stellar coordinates. Once the
coordinate list is compiled, the \texttt{psfmeasure} routine  analyzes the stars
at those coordinates and returns a list of FWHMs. An average FWHM is calculated
from this list. The relevant photometric constant  (magnitude scale zero point) 
was taken from the WFC3 manual\footnote{ {\tt
http://www.stsci.edu/hst/wfc3/phot\_zp\_lbn}, the version ``Prior to  March 6,
2012'' } for Vegamags.  They are then entered into the photometry parameters
file (\texttt{photpars}) and the average FWHM and background standard deviation
are entered into the data parameters file (\texttt{datapars}). An aperture
radius of  10 pixels (0$\farcs$4) was used.  The IRAF photometry package
(\texttt{phot}) is then run on the image to perform aperture photometry,  using
the star coordinates and parameter files as input to produce the stellar
magnitudes and their errors.

Examples of the WFC3 images in the F845M filter are provided for R Cru
(Fig.~\ref{rcru.im}), V~Cen (Fig.~\ref{vcen.im}), and FF~Aql 
(Fig.~\ref{ffaql.im}).  

\section{Finding Resolved Physical Companions}

\subsection{Isochrones}

The goal of our project is to identify candidate resolved physical companions of
the Cepheids imaged in the survey. The selection criterion is that the candidate
companion star must lie in the color-magnitude diagram (CMD) near an isochrone
for the typical age of a Cepheid (50~Myr), corrected to the distance and
reddening of each Cepheid. 

Isochrones in the WFC3 filters that we used, and in the ground-based $V$ and
Kron-Cousins $I$ bands, are available from two sources. We created a 50~Myr
isochrone as follows. (1)~For the unevolved lower main sequence (defined here as
$V-I>0.75$), we used the Dartmouth Stellar Evolution Database\footnote{Dotter et
al.\ (2008); data tables are at {\tt http://stellar.dartmouth.edu\slash
models\slash index.html}, retrieved 2014 June 9}. This compilation gives
isochrones in the WFC3 F621M and F845M filters, and in $V$ and $I$\null. We
selected isochrones for solar metallicity ($\rm[Fe/H]=0$). Unfortunately, for
the lowest available age in the Dartmouth isochrones, 1~Gyr, there is a large
gap in coverage from $V-I\simeq0.9$ to 1.5. We therefore used data from
the 5~Gyr isochrone, which is free of large gaps and agrees very well with the
1~Gyr isochrone in the color ranges where they do overlap. (2)~For the stars
with $V-I<0.75$, which have begun to evolve off the zero-age main sequence in
the available Dartmouth isochrones, we used a ``Padova'' 50~Myr
isochrone\footnote{Bressan et al.\ (2012); data tables are at {\tt
http://stev.oapd.inaf.it/cmd}, retrieved 2014 June 11} for a heavy-element
content of $Z=0.0152$. We again obtained isochrones in WFC3 F621M and F845M, and
in ground-based $V$ and $I$.

For the unevolved main sequence, we note that the Dartmouth and Padova
isochrones agree reasonably well from $V-I\simeq0.75$ to about 1.25. But then
they begin to diverge as we move further down the main sequence, with the Padova
absolute magnitudes becoming progressively fainter than the Dartmouth values. At
$V-I\simeq2$, the Padova $M_V$ values are fainter than Dartmouth by about
0.6~mag. By comparison with empirical absolute magnitudes for
lower-main-sequence stars with accurate parallaxes (e.g., those assembled by
E.~Mamajek\footnote{Pecaut \& Mamajek (2013); data tables are at {\tt 
http:/\slash www.pas.rochester.edu/$^\sim$emamajek\slash
EEM\_dwarf\_UBVIJHK\_colors\_Teff.txt}}), we find that the Dartmouth isochrones,
in $V,(V-I)$, give better agreement with real stars. We therefore adopted a
combination of the Padova 50~Myr isochrone for $V-I<0.75$ (and the corresponding
isochrone for $F621M$ and $F845M$ magnitudes), and the Dartmouth 5~Gyr isochrone
for the unevolved cooler main-sequence stars, as our standard Cepheid-age
solar-composition isochrone.

In the discussion below, we will find it useful to transform the WFC3 $F621M$
magnitude and $F621M-F845M$ color index to ground-based $V$ and $V-I$\null. A
least-squares polynomial fit to the Padova-Dartmouth tables gives the following
relations, which have residuals no larger than $\pm$0.02~mag:
\begin{eqnarray}
V & = & F621M + 0.115637 x^7 - 0.98113 x^6 + 3.09425 x^5 
  - 4.26033 x^4 + 2.03240 x^3 \nonumber \\ 
  &   & + 0.14579 x^2 + 0.46920 x + 0.041 \nonumber \\
\noalign{\noindent and}
V-I & = & -0.06353 x^5 + 0.48892 x^4 - 1.20278 x^3 
  + 0.86796 x^2 + 1.36607 x + 0.002 \, , \nonumber 
\end{eqnarray}
where $x=F621M-F845M$.

In order to create an isochrone appropriate for each individual Cepheid, we need
to correct it to the reddening and distance of the Cepheid, given in columns~4
and 5 of Table~1, respectively. To obtain the reddening law in the WFC3 F621M
and F845M filters, we  used the formulae of Cardelli, Clayton, \& Mathis (1989),
with effective wavelengths of 6210 and 8450~\AA, respectively. This yields the
relations $E(F621M-F845M) = 1.2\, E(B-V)$ and $A_{F621M} = 2.9E(B-V)$.



Examples of comparisons of the WFC3 photometry with the 50~Myr isochrone are
given for the fields surrounding R~Cru (Fig.~\ref{rcru.cmd}) and V~Cen
(Fig.~\ref{vcen.cmd}).  In these figures we have highlighted
the region in the isochrone where we
expect to be able to detect X-ray emission from young, low-mass stars (i.e.,
those with spectral types of F2~V through K7~V, with unreddened $F621M-F845M$
colors ranging from 0.27 to 1.06). Young M dwarfs also produce
X-rays, but the flux drops  quickly with advancing spectral type, and at the
distances of our Cepheids we do not expect X-ray emission to be detectable
(Evans  et al.\ 2016). We have therefore neglected candidate M~dwarf companions,
even though they would be easily detected in  the WFC3 photometry.

\subsection{Candidate Companions} 

\begin{figure}
 \includegraphics[width=\columnwidth]{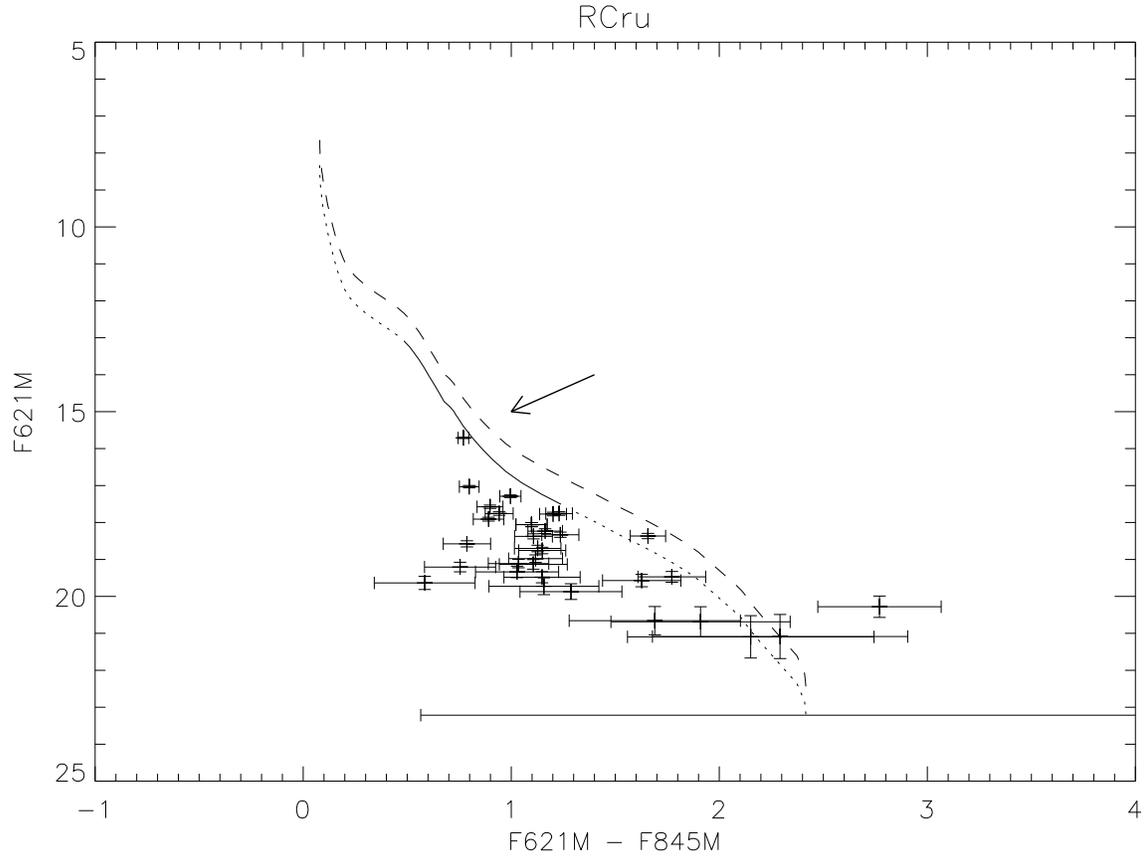}
\caption{The CMD for R Cru. Data points are from the stars detected in
the F621M and F845M  
images.  The  solid line is the  ZAMS from F2  to K7; the dotted line
extends it  outside this range; the dashed is the ZAMS F621M - 0.75 mag to account for binaries.  
The arrow indicates the possible companion. \label{rcru.cmd}}
\end{figure}

\begin{figure}
\includegraphics[width=\columnwidth]{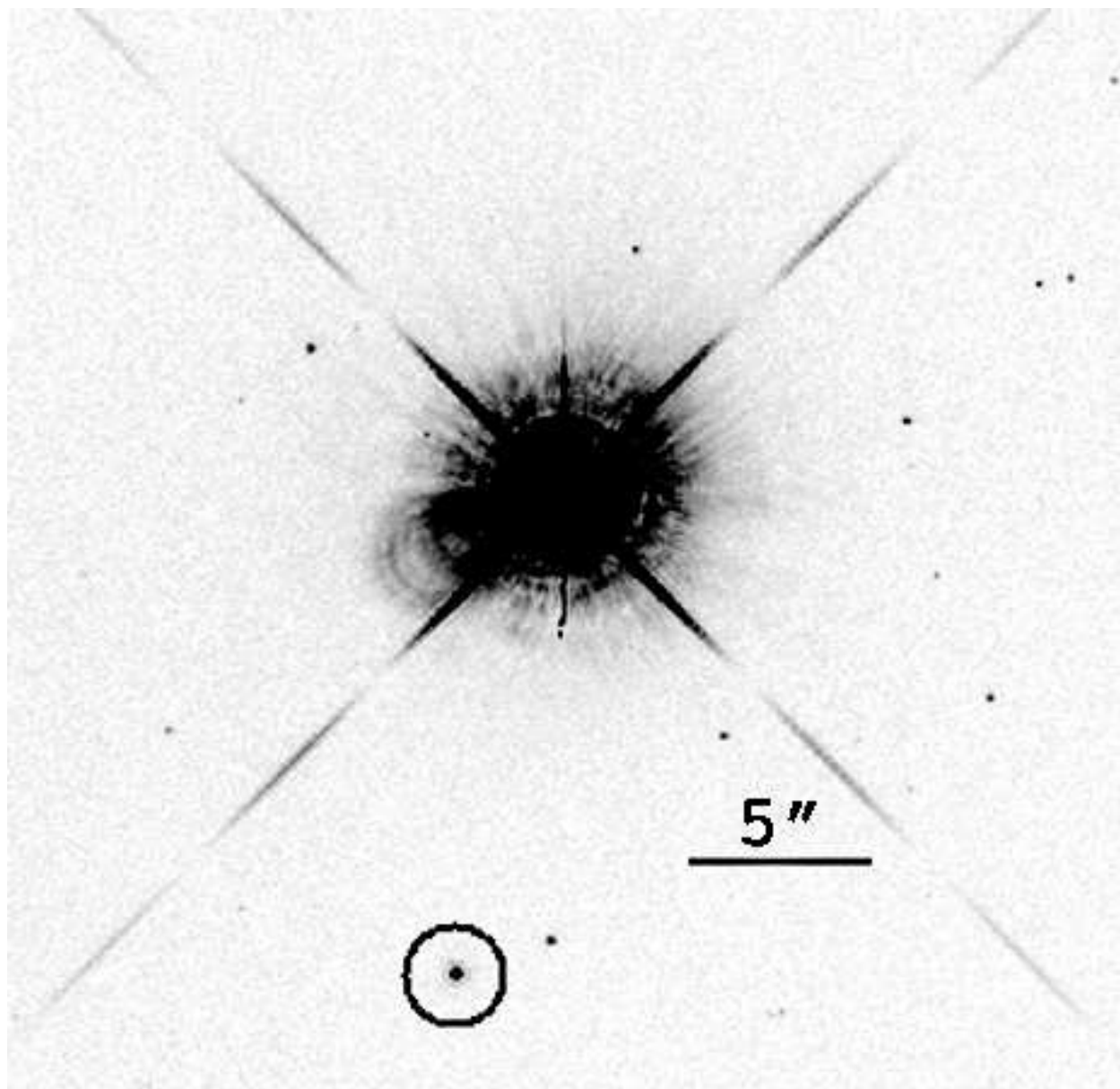}
\caption{V Cen image in the  F845M filter. Symbols are the same 
as in Fig. ~\ref{rcru.im}   \label{vcen.im}}
\end{figure}

\begin{figure}
 \includegraphics[width=\columnwidth]{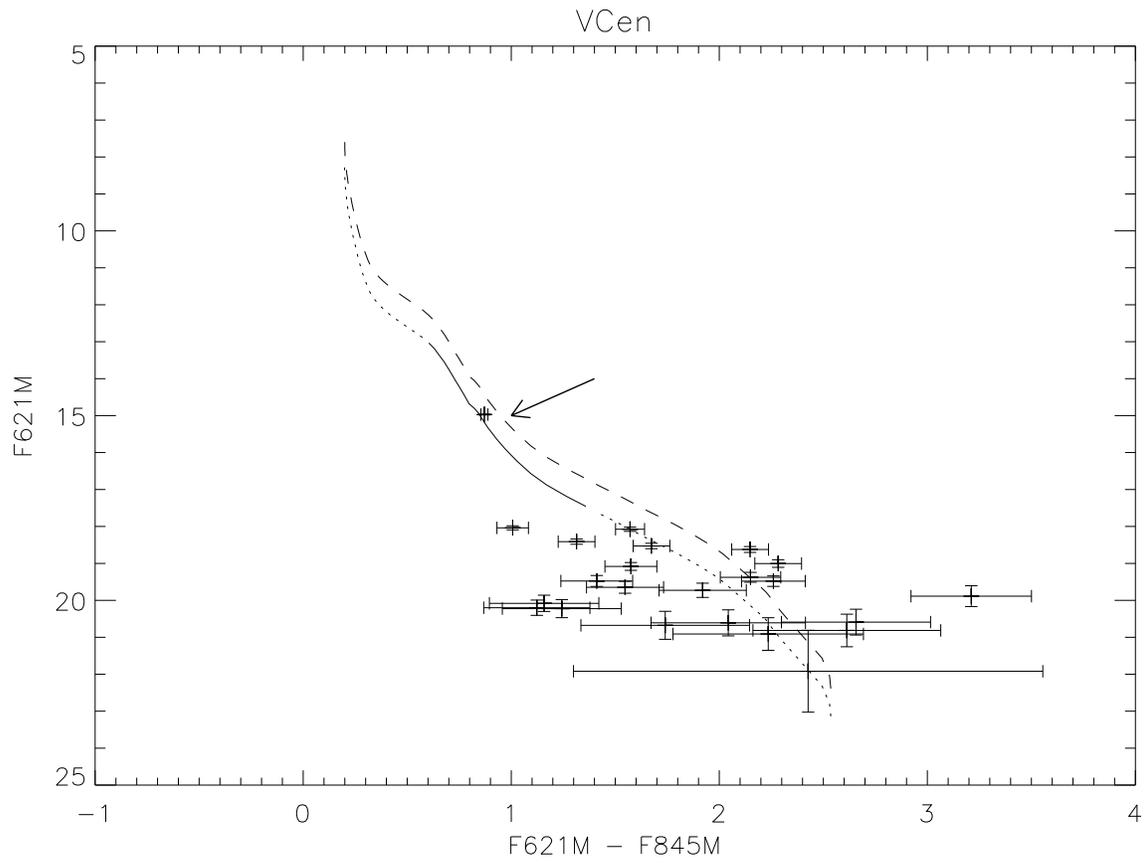}
\caption{The CMD for V Cen.   Symbols are the same as for Fig.~\ref{rcru.cmd}  \label{vcen.cmd}}
\end{figure}

\begin{figure}
 \includegraphics[width=\columnwidth]{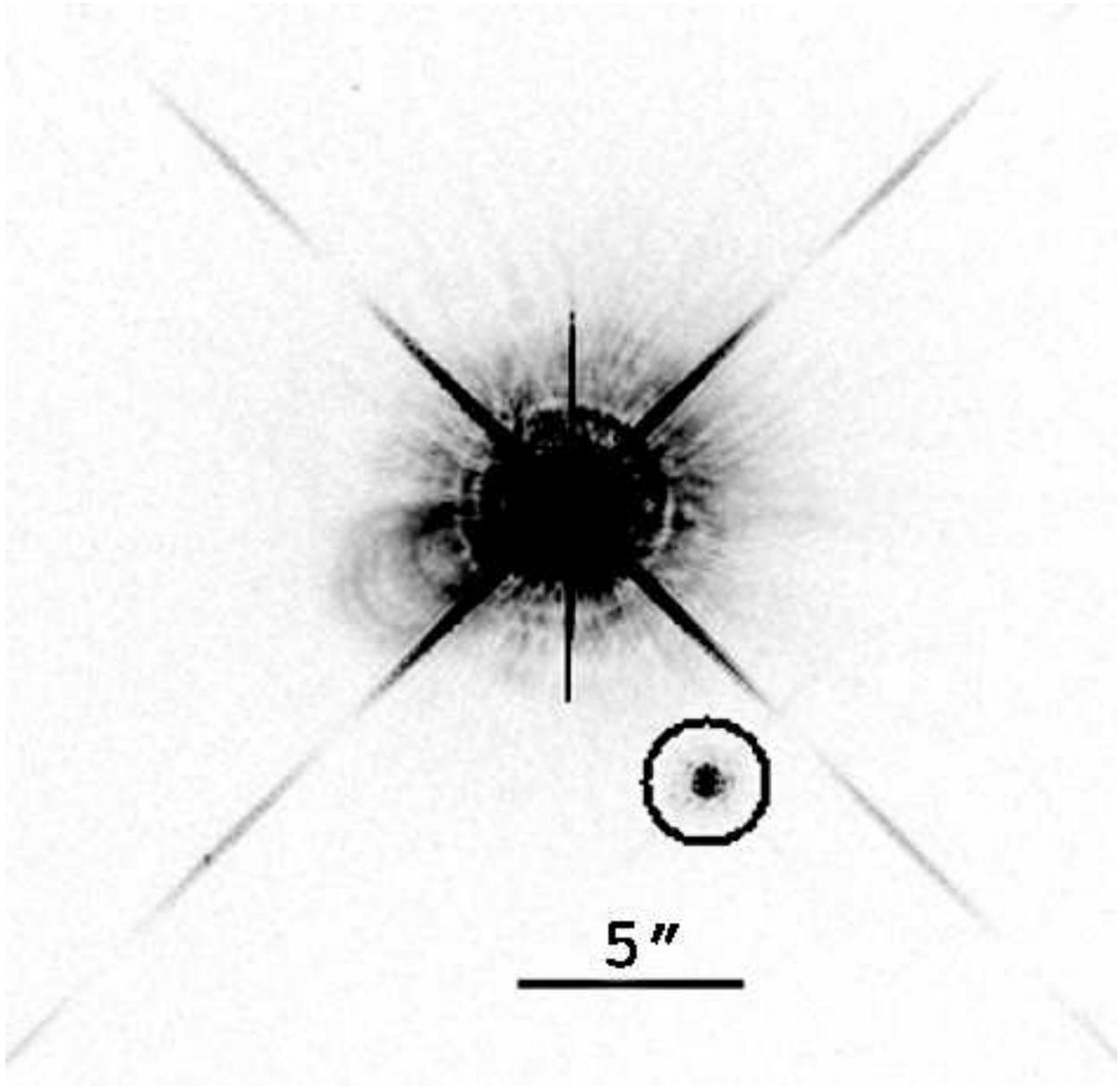}
\caption{FF Aql image in the F845M filter.  Symbols are the same as in    
Fig. ~\ref{rcru.im}  \label{ffaql.im}}
\end{figure}

Using the \HST\/ $F621M$ vs.\ $F621M - F845M$ CMDs, we selected candidate
physical companions in the following way. In addition to the ZAMS, a line was
placed 0.75 mag above it, indicating the upper limit of binaries of identical
mass (e.g., Fig.~\ref{rcru.cmd}). The region between these lines is the region
where companions are expected, and a list was generated (Appendix A) of stars
within 2$\sigma$ of this region. The identification criteria were guided by
experience with the low-mass members of the S Mus cluster identified on an  {\it
XMM\/} image (Evans et al.\ 2014). Occasionally additional judgment was invoked.
For instance, fainter stars with errors in the colors greater than 0.2 mag were
dismissed as too uncertain for further consideration. 

We examined the CMDs for all the Cepheids for possible companions, finding
candidates for 39 out of the 70 targets. These 39 Cepheids, and their candidate
companions, are listed in Table~A1 in Appendix A\null. Table~A1 contains all
candidate companions lying within 2$\sigma$ of the ZAMS in the CMDs, and hotter
than spectral type M0. The first two columns list the $F621M$ and $F621M -
F845M$ magnitudes and colors, and their errors. For convenience, the next two
columns provide the transformed $V$ and $V-I$. The following two columns give
the angular separations and position angles. The final column provides the
projected separations converted to AU, using the distances from
Table~\ref{survey}. 

Table A1 lists possible companions $\geq$5$\arcsec$ from the Cepheid, which is
the sample discussed in the rest of this paper. Additional possible companions
were identified by our search technique, lying less than 5$\arcsec$ from the
Cepheid. These are listed in Table~A2, but they will only be discussed in
Paper~III, together with an additional and more complete catalog of candidate
companions identified in PSF-subtracted images. 

Figures~\ref{rcru.cmd} and~\ref{vcen.cmd} show two examples of
the $F621M$ and $F621M - F845M$ 
CMDs, for R~Cru and V~Cen, respectively.  
For these two Cepheids,  there is one star within 2$\sigma$ of this part of the
ZAMS, which are the potential companion candidates.  These stars are circled on
Figs.~\ref{rcru.im} and \ref{vcen.im}. These figures are typical in two
respects.  The CMDs rule out all but a very few stars in the field as possible
companions.  That is, the companion range we were working with was something of
a ``sweet spot,'' with relatively few interlopers.  In addition, even at the
low galactic latitudes of Cepheids, the field only becomes heavily populated
fainter than the M0 limit.  





\section{X-ray Confirmation}



Because faint red stars are common in the Galactic plane, it is important to
confirm that our candidate companions are physical, rather than chance
alignments with field stars. As discussed above, chance alignments would be much
more common in a deeper survey, but it is still important to evaluate the
relatively bright candidates in Table~A1. Physical companions of Cepheids must
be the same age as the Cepheids themselves, typically 50 Myr. Low-mass stars
with a chromosphere have decreasing rotation rates as they age because of
magnetic braking (e.g., Pallavicini et al.\ 1981). For this reason, stars young
enough to be Cepheid companions are easily distinguished from old field stars by
their X-ray fluxes. We are conducting a series of observations of the Cepheids
in Table~A1 using the {\it XMM-Newton} satellite, to be described in Paper~IV
(Evans et al.\ 2016). An example is provided by S~Mus (Evans et al.\ 2014). So
far 14 of our 39 Cepheids with possible resolved companions have been observed
with {\it XMM\/} ($\ell$~Car, V659 Cen, V737 Cen, R Cru, S Cru, X Cyg,  V473
Lyr, R Mus, S Mus, S Nor, Y Oph, V440 Per, U Sgr, and Y Sgr), to a depth where
essentially all the K stars (and hotter) at the distance of the Cepheid with an
age of 50 Myr would be detected. There is also an upper limit on W~Sgr from the
{\it ROSAT\/} All Sky Survey (\S\ref{wsgr} below). 

The X-ray results are summarized as follows:

$\bullet$ Our wide candidate companions are overwhelmingly {\it not\/} young
X-ray active  stars, and hence are unlikely to be physical companions.  
Twenty-three  candidate companions  with separations $\geq$8$\arcsec$  (with the
exception of S~Nor; see below), and three with smaller separations (V737 Cen, R
Mus, and Y Sgr), were rejected because of the lack of an X-ray detection.




$\bullet$ {\bf R Cru} has an X-ray source in its field; however, at {\it XMM\/}
resolution, the source could be assigned either to the Cepheid or the candidate
companion star at 7$\farcs$6 (6,330~AU). However, there is an even {\it
closer\/} source in the WFC3 images, to be discussed in Paper~III because its
separation is only $1\farcs9$. In the following we assume that this closer
source produces the X-rays. 

%

$\bullet$ {\bf S Mus} similarly has an X-ray source, but again {\it XMM\/}
observations cannot distinguish between the Cepheid and the possible companion
at 5$\farcs$0 (3,950 AU). We have subsequently obtained a {\it Chandra} observation
of the S Mus field.  Because of its higher spatial resolution, we can 
now conclude that the X-rays come not from the  5$\farcs$0 companion, but 
from the Cepheid/spectroscopic binary.  (Full discussion of the results 
is in preparation.)  Thus at present we can say that there are no 
resolved physical companions of Cepheids at 4,000 AU or wider.

$\bullet$ {\bf S Nor} is our one Cepheid with an X-ray source unequivocally at
the position of a resolved companion, with a projected separation of 14$\farcs$6
(13,300 AU). This is significantly wider than the companions of either R Cru or
S Mus. However, S~Nor is a member of a populous cluster. For this  reason, and
based on the discussion below in \S\ref{field.dens}, we consider it likely that
it is a chance alignment with a cluster member, rather than a gravitationally
bound binary companion.  

Thus our preliminary ``working model'' is that physical companions of Cepheids
are found at separations within about 4,000 AU; however, in a cluster there
may be wider alignments, but they are probably not bound system members.




\subsection{The Most Probable Companions} 













One of the goals of this study is to examine the range of separations of
physical companions to Cepheids. We use the X-ray results on the lack of wide
physical companions to apply another filter to the list of possible companions
(Table A1).  Specifically, we examine in detail the Cepheids with candidates
lying at separations $\geq$ $5''$ and $\leq$ 6330 AU. The 6330 AU upper limit is
based on R Cru, the widest of the {\it possible\/} X-ray companions,
although we conclude in Paper IV that the X-rays are most probably 
produced by a closer companion (to be discussed in Paper III).
Table~\ref{prob} lists this subset of 11 candidate companions (of ten Cepheids).
The first five columns are taken from Table~A1. Column 6 in Table~\ref{prob}
summarizes the X-ray results. 


We now examine further evidence whether the stars in Table~\ref{prob} are
physically related to the Cepheids. Four of the Cepheids have been observed in
X-rays (V737 Cen, R Mus, W Sgr, and Y Sgr), and were not detected; these four
are listed at the bottom of the table; see \S\ref{wsgr} for W Sgr). Thus, for
the total sample of 70 Cepheids in our survey, ten of them have potential
resolved companions ($\geq$ $5''$)  within 7000 AU.
However, when the
X-ray results are folded in, four are eliminated. Of the six stars observed by
{\it XMM\/} or {\it ROSAT\/}  (including R Cru and S Mus), four were not
detected, and hence are probably field stars. The remaining two sources that
were detected (R Cru and S Mus) are so close to the Cepheid that both the
Cepheid and the companion are within the {\it XMM\/} PSF\null.
However, in both cases, additional evidence indicated that the resolved 
companion candidate is not the X-ray source.
R Cru has a
closer companion which we consider a more probable X-ray candidate (see Paper III).
The {\it Chandra} observation of S Mus also shows that the X-rays are produced
by the Cepheid/spectroscopic binary. 

We can use Table~\ref{prob} to make an estimate of the 
companion frequency ($\geq$ $5''$) within 7000 AU.
from the Cepheid.
 Since none of the 6 possible companions $\geq$ $5''$ which have 
observed in  X-rays have been detected, the  remaining four Cepheids 
(TT Aql, FF Aql, AP Sgr, and RV Sco) might have the same detection rate.
However, the companions of FF Aql (\S\ref{ffaql}) and RV Sco (below) are both
likely to be physical companions since they appear to share a proper motion. 
Furthermore, anticipating the discussion
for field density (\S\ref{field.dens}), the fact that TT Aql has 2 possible
companions as well as 5 possible companions in the whole field makes it unlikely  
that the 2 stars listed in Table~\ref{prob} are bound binary companions.  This leaves 
only 1 star (AP Sgr) in Table~\ref{prob} as a possiblity, but with the low X-ray 
detection rate, it is also unlikely.  
Thus, the binary rate in this separation range is 2 (RV Sco and FF Aql) out of 
70 or  3\%.

  

 {\bf RV Sco} has additional information from proper
motions.  The Washington
Double Star  Catalog (WDS)\footnote{\tt
http://www.usno.navy.mil/USNO/astrometry/optical-IR-prod/wds/WDS} has measures
for the pair in 1925 and 1987, showing no  relative motion between the two
stars.  The total motion of the Cepheid is about half an arcsecond in that
period, indicating that the two are moving together, and thus likely bound. 


 Column 9 in Table~\ref{prob} lists the intrinsic $(V-I)_0$ color, using
the  $E(B-V)$ from Table~1 and  $E(V-I)/E(B-V) = 1.15$ (assuming, of course,
that the stars have the same reddening as the Cepheids). All the possible
companions have the colors of K stars, except for FF Aql and RV Sco, where the 
possible companions have  hotter colors.  (These colors are hot enough that 
the companions would not be expected produce X-rays.)
K stars are certainly reasonable
companions for Cepheids. On the other hand, they are also the most plentiful
field stars in the range under  consideration, and hence the most likely to be
chance alignments.  The fact that  the two   most probable companions 
(FF Aql and RV Sco) in Table~\ref{prob} are more massive than K stars
suggests the companion distribution is top heavy
as compared to the Initial Mass Function (IMF).  This
is clearly not due to the
detection limit, since K dwarfs are easily detected in the WFC3 survey.      

Anticipating the discussion of field density (\S\ref{field.dens}), we list in
Column 8 the  number of possible companions (Table A1) with separations
$\geq$5$\arcsec$ for each Cepheid.   As fully discussed below
(\S\ref{field.dens}),  a large number of possible companions increases the
probability that the possible companions are not gravitationally bound. On these
grounds, the companions of TT~Aql and RV Sco are suspect.   We note that the
discussion of proper motions above implies that the RV Sco companion is  indeed
related to the Cepheid.  

 The following summarizes the discussion above of 
  the frequency of resolved companions as well at the 
outer extent: 

$\bullet$ Both of the estimates of the companion frequency for separations 
$\geq$5$\arcsec$ are 3\% or less. 

$\bullet$  We have identified no probable companions
wider than  RV Sco (4520 AU).  Thus 4000 AU is a 
reasonable estimate for the extent of companions from our summary.





\section{The Nearer Cepheids}

In this section and the next one, 
we discuss two tests to confirm the 
relative scarcity of physical companions with a projected separation from
the Cepheid of $\geq$5$\arcsec$.

In order to confirm the X-ray results that stars in the outer parts of the
field  are chance alignments, we created Table~\ref{near},
containing the subset of Cepheids nearer than 600 pc. These are the Cepheids
that are least likely to  be  contaminated by field stars at the brightnesses
expected for physical companions.    Column 2 indicates
whether there is a possible companion on the WFC3  images with ``Yes''; column 3
indicates whether the possible companion has   {\it not\/} been detected in an
X-ray observation with a ``No.''    Since the nearest Cepheids have the most
complete information,  we are able to discuss three of them in more detail, as
follows.

\subsection{$\delta$ Cep}

$\delta$ Cep  poses a special challenge in determining its relation to the star
HD~213307, a late B star 40$\arcsec$ away.  Parallaxes were determined for both
stars using the \HST\/ FGS by Benedict et al.\ (2002).  They found both stars to
be at the same distance  within the errors, and also found that HD 213307 is
itself a binary from the astrometry.  Proper  motions and radial velocities  of
$\delta$ Cep and  HD 213307 are similar but not identical. Both stars are listed
as members of the newly discovered Cep OB6 association (de Zeeuw et al.\ 1999). 
HD 213307 is not within our WFC3 image, but that is  because $\delta$  Cep is so
nearby.  At its distance, the projected  separation is 10,200 AU\null.  This is
larger than the separations (Table~\ref{prob}) for the most probable companions,
set largely by the X-ray  results.  It is also the outlier in the separation
distribution for Cepheids with  reasonably massive companions (Fig.~5 in Evans
et al.\ 2013).  Is HD 213307 gravitationally bound to $\delta$~Cep?  If so, the
separation is unusually wide.  On the other hand,  because they belong to an
association, the possibility of a chance alignment with a star at the same
distance is enhanced. 


\subsection{FF Aql} 
\label{ffaql}

\begin{figure}
 \includegraphics[width=\columnwidth]{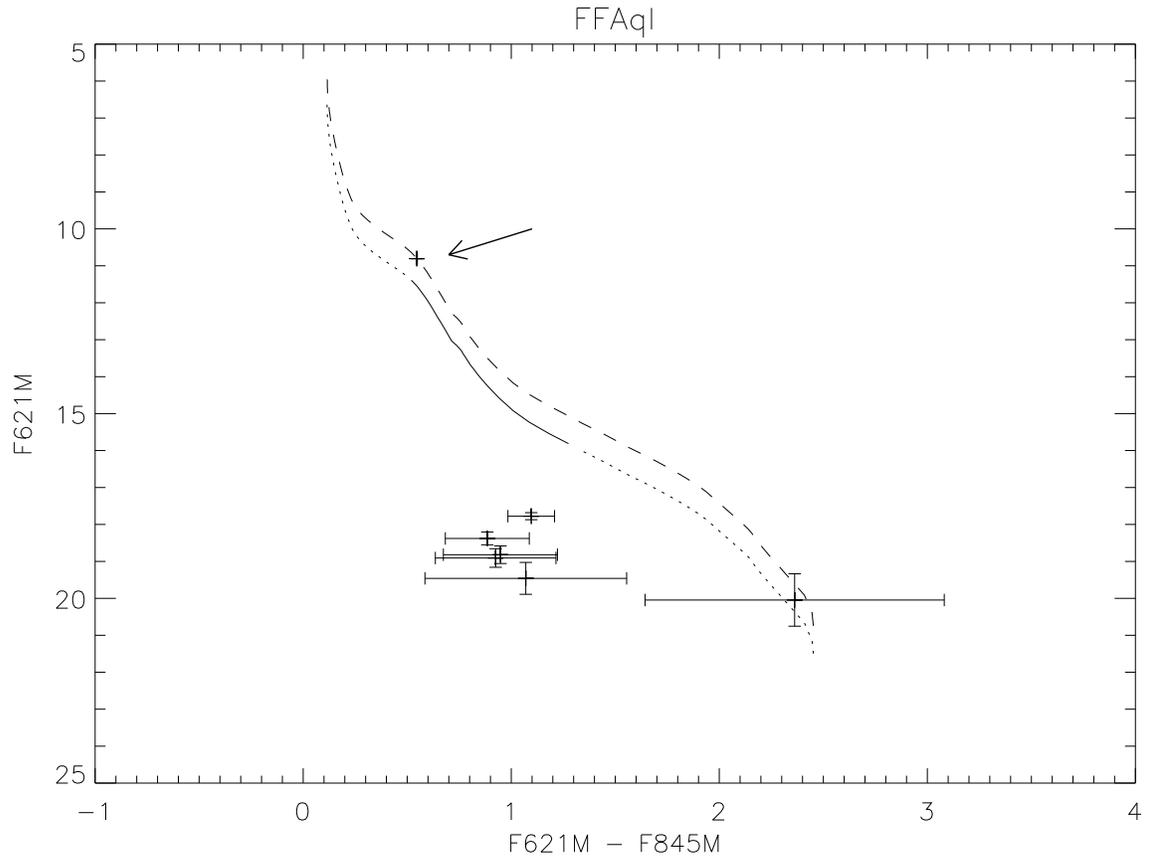}
\caption{The CMD for FF Aql.  The symbols are the same as Fig.~\ref{rcru.cmd}.  \label{ffaql.cmd}}
\end{figure}

FF Aql has been a candidate for a wide binary for many years. It has a
comparatively bright possible companion 7$\arcsec$ from the Cepheid
(Fig.~\ref{ffaql.im}).  This small separation, however, has made it  difficult
to determine the colors of the companion from the ground, because of scattered
light from the Cepheid.  Udalski \& Evans (1993) concluded that  the two stars
are not related based on photometry.  The \HST\/ results  (Table~\ref{prob} and
Fig.~\ref{ffaql.cmd}) place the companion within the  expected main-sequence
band.   Thus, based on the CMD and separation, the companion to FF Aql is  probably
a physical companion.  The $V-I$ color in Table~\ref{prob} (dereddened)
corresponds to a main-sequence star between F5 and G0 (Drilling \&  Landolt
2000).  This would make the Cepheid a member of a triple system, since it is
also a member of a spectroscopic binary.  
As with RV Sco, measurements of the  positions of the two stars since 
1886 (WDS) indicate no relative motion, indicating that they are a physical
pair.

\subsection{W Sgr}
\label{wsgr}

W Sgr is already known to be a member of a complicated multiple system.  
It is a spectroscopic binary (Petterson et al.\ 2004, and references
therein) with a very low-amplitude orbit, but 
a fairly massive companion (Evans 1995).  Recently Evans, Massa, \& Proffitt 
(2009) used an \HST\/ 
observation to demonstrate that the hottest star in the system is actually a
resolved companion (projected separation 0$\farcs$16), 
not the secondary in the spectroscopic binary.  
Thus, the star with a separation of
$6\farcs3$ (2580 AU) in Table~2 would be the fourth star in the system
if it is a physical companion.

Because W Sgr is a nearby system, the results of the {\it ROSAT\/} All Sky 
Survey\footnote {http://www.xray.mpe.mpg.de/cgi-bin/rosat/rosat-survey} provide
useful information.   No source was visible at the position of the system. From
the background and exposure map, an upper limit to the count rate of 0.0012 cts
sec$^{-1}$ was derived. (The PSF of {\it ROSAT\/} is $\sim$45$\arcsec$.)  
The {\it ROSAT\/} PSPC (energy 0.1 to
2.4 keV) was converted to a flux in 0.5 to 8.0 keV, using  PIMMS with
$kT = 0.48$ keV and 
$N_H=  10^{21}$ cm derived from the $E(B-V)$.  The flux upper limit derived from
the count rate is $1.61 \times 10^{-14}$ ergs cm$^{-2}$ sec$^{-1}$.  At a
distance of 409 pc, this  becomes an upper limit of $L_X = 3.2 \times 10^{29}$
ergs sec$^{-1}$   ($\log L_X = 29.51$).  This is slightly above the limit
for X-ray  flux, $\log L_X = 29.2$,  used in Paper~IV  (Evans et al.\ 2016) as
the lower limit for low-mass stars young enough to be Cepheid companions.
However, it is low enough that we consider it improbable that it is  actually a
physical companion, and treat it as such here.

\subsection{Polaris}
\label{pol}

We add one additional star to the discussion of nearby Cepheids, even though it was not
included in the WFC3 survey: Polaris.  Like $\delta$ Cep, it has a resolved 
companion 18$\arcsec$ from the Cepheid.  The companion Polaris B has 
a proper motion consistent with  orbital motion in a wide orbit.
The spectral type of the companion  (summarized by Evans et al.\ 2010) is 
F3~V\null.  A {\it Chandra\/} observation found that it is not an X-ray source  
(Evans et al.\ 2010), which is not surprising for an early F star.  
Using the distance from {\it Hipparcos\/} (130 pc; Feast \& Catchpole 1997;
van Leeuwen 2007), the companion has a projected separation from the 
Cepheid of 2340 AU, well within the range of the most probable companions
(Table~\ref{prob}). To complete the list of system components, Polaris is a
member of a 30 year astrometric binary, whose secondary has been 
detected by {\it HST} (Evans et al. 2008).


\subsection{The Nearer Cepheids: Summary}

The sample of the nearest Cepheids in Table~\ref{near} has 21 stars, of which
seven have possible companions in the WFC3 images. An additional nearby Cepheid,
$\delta$ Cep, has a companion which is not on a WFC3 image because of its wide
angular separation.  Of the eight stars with possible companions, five have been
observed with {\it XMM\/} or {\it ROSAT}, and the companions were not detected;
hence they are classed as chance optical alignments with field stars. The three
remaining possibilities  in Table~\ref{near} are $\delta$ Cep, FF Aql, and V636
Cas.  If we further apply the criterion from the {\it XMM} observations in the
previous section that only stars closer than 6330 AU are probable physical
companions, V636 Cas and $\delta$ Cep are disqualified. This leaves one star out
of 21 in Table~\ref{near} with a probable resolved companion (5\%). This is, of
course, a very small sample, but since this subsection of the sample is the
least likely to be contaminated by field stars, it confirms the small companion
fraction from the combined WFC3 and {\it XMM\/} observations. (Including
Polaris, the companion fraction rises to 2 out of 22 [9\%].)

We note also that the companion of Polaris is significantly more massive
than a K star, like those of FF Aql and RV Sco. On the other hand, the four stars 
rejected by X-ray observations  Table 2 are all K stars, leaving two
Cepheids with possible K companions (TT Aql and AP Sgr). This suggests 
a preference for more massive companions than the distribution of the IMF. 

In our working model (\S4), we add $\delta$ Cep to S Nor \#4 as Cepheids which 
are members of known clusters or associations, and hence likely to have a chance alignment
with a related but not gravitationally bound star.

\section{Field Stellar Density}
\label{field.dens}

\begin{figure}
 \includegraphics[width=\columnwidth]{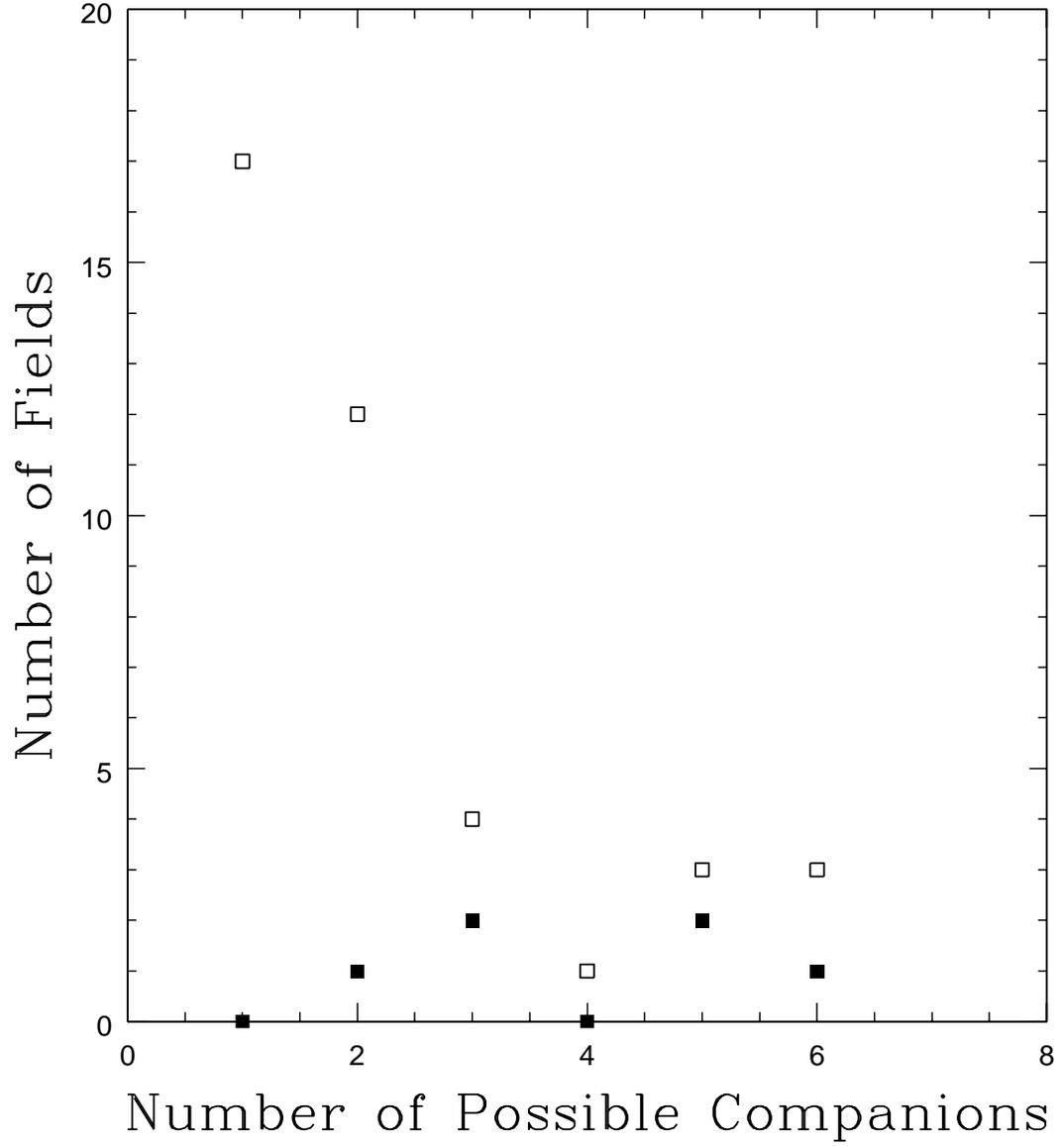}
\caption{The number of fields for a given number of possible 
companions.  Solid squares: fields with a possible companion $<$5$\arcsec$;
open squares:  fields with a possible companion $>$5$\arcsec$.
\label{dens}}
\end{figure}

The second test we have made is to check whether the occurrence of a possible
companion depends on the surface density of stars in the field. We generated a
frequency distribution for the number of fields with 1, 2, 3, $\dots$ companions
from Table~A1 (i.e., the companions $>$5$\arcsec$ from the Cepheid), and
similarly for Table~A2 (companions $<$5$\arcsec$ from the Cepheid).
Fig.~\ref{dens} shows the comparison. Only a very small fraction of the fields
with one or two companions have possible companions $<$5$\arcsec$. In contrast,
approximately half the more dense fields (three or more possible companions)
have possible companions closer than 5$\arcsec$. This is consistent with the
possibility that denser fields are linked with increased chance alignment. This
supports the working model that some Cepheids may have been formed in a loose
grouping as well as gravitationally bound systems. Fig.~\ref{dens} confirms
that, in the overwhelming majority of fields with 0, 1, or 2 possible companions
(59 fields), the occurrence of a possible companion $\geq$5$\arcsec$ is very
rare (2\%).



\section{Discussion}

We have made a further check on the working model, specifically that in addition
to gravitationally bound system members, star formation may have produced some
surrounding groups with a range of densities from very low to well-populated 
well-known clusters. Two recent studies have  identified Cepheids related to
clusters based on criteria including proper  motions (Anderson et al.\ 2013;
Chen et al.\ 2015). Of the Cepheids that they conclude are definitely associated
with clusters or associations, seven are in our survey (U Sgr, SU Cyg, S Nor, BB
Sgr, V Cen, S Mus, and X Cyg---although X Cyg is only considered a definite
member by Chen et al.). For these stars we examine the evidence of the existence
of nearby stars at the same distance using Tables A1 and A2. None of these
Cepheids are in the 30 fields (43\%)  which have no possible companions. Three
(U Sgr, S Nor, and X Cyg) have more than one possible companion. Of the 4 which
have been observed in X-rays (U Sgr, S Nor, X Cyg, and S Mus), one 
X-ray source at a wide companion (S Nor). 
Thus, even for this group there is at only one
possible related star outside the separation range (4000 AU) where we
find gravitationally bound binaries. We draw attention to possible wide
associations, particularly as low density groupings may be of particular
interest when {\it Gaia\/} results are available. 

This paper focuses on the extent of gravitationally bound systems for $\sim\!6\,
M_\odot$ Cepheids. The fraction with companions wider than $10^3$~AU is very
small confirming that the frequency distribution for binary systems 
peaks at much smaller
separations. Although full analysis of the frequency distribution as a function
of separation/period awaits the discussion of closer companions (Paper~III), we
can compare results on the extent of companions with studies in other mass
ranges. For O stars, the recent high-resolution study of Sana et al.\ (2014;
SMASH) found a decrease in the number of comparatively bright companions at
separations of approximately 2000 AU, as discussed by Evans et al.\ (2016;
Paper~IV). For solar-mass stars, Raghavan et al.\ (2010) and Tokovinin (2014)
both find a decrease in the frequency of companions at periods $>$10$^5$ years
($\simeq$3000 AU), also in approximate agreement with the  results of the
present survey.

\section{Summary}


We report here the results of an \HST\/ WFC3 snapshot imaging survey of 70
classical  Cepheids.  This paper (Paper~II in the series) discusses possible
companions  with separations $\geq$5$\arcsec$ from the Cepheid.  

$\bullet$ We identify possible 
companions by comparison of the $F621M$ vs.\ $F621M-F845M$ CMD with evolutionary
tracks at the distance and with the reddening of the Cepheids with a width
allowing for a binary sequence.  The list of 39 possible 
Cepheids with companions (Table A1) 
should fully cover  the spatial extent of possible companions and  identify them
through main-sequence stars of K spectral types.  

$\bullet$ Fourteen of the possible companions have been observed with {\it
XMM\/} (details in Paper~IV) to distinguish active  stars as young as Cepheids
from old field stars. From these observations,  we find no young stars at a
larger separation from the Cepheid than that of S Mus  (5$\farcs$0 or 3950 AU). 
However, the {\it XMM\/} observation does not resolve the  companion and the
Cepheid. A {\it Chandra} observation shows that the X-rays are not produced
by the star at  5$\farcs$0 separation, but from the Cepheid/spectroscopic binary.

$\bullet$ Based on the X-ray results (Paper~IV) of a subset of 14 Cepheids with possible
companions we estimate a frequency of companions $\geq$ 5$\arcsec$ to be 3\% or less.  

$\bullet$ Companions more massive than K stars predominate among the most probable 
companions, which cannot be due to the WFC3 detection limit. 

$\bullet$ We have confirmed the outer extent of gravitationally bound systems from the subset
of  Cepheids closer than 600 pc, which are the least susceptible to chance alignments with
field stars, and find a comparable frequency of probable companions.  

$\bullet$ Similarly using the list of  companions closer than 5$\arcsec$ of the 
Cepheids  (Table A2, the most likely to be bound system members), we  compare the number 
of close possible companions with the field density as indicated by the 
number of possible companions in Table A1.  Fields with three or more possible 
companions are more likely to have close companions, which we attribute to 
increased probability of chance alignment in the denser fields (loose groupings).  

$\bullet$ The working model is  
that  the possible companions stars in 
Table A1 may include  both bound system members, and also  stars formed at the same time in
the same neighborhood.  This may  be the case of HD 213307 ($\delta$ Cep companion) and
S Nor \#4, since both the Cepheids are members of clusters or associations, which  may 
account for their unusually wide separations.

Thus, in this paper we focus on the outer extent of Cepheid multiple systems. 
As noted in the introduction, systems this wide could easily support an inner
binary,  hence be part of a triple system.  In the next paper (Paper~III) we
will discuss companions less that $5''$ from the Cepheid and then the
characteristics of the combined population of resolved binary companions.











\acknowledgments

Financial support from STScI grants GO-12215.01-A and GO-13368.01-A
and also Chandra X-Ray Center NASA Contract NAS8-03060   
is gratefully acknowledged.
Vizier and  SIMBAD were used in the preparation of this study.
This research has made use of the Washington Double Star Catalog maintained 
at the U.S. Naval Observatory.




\appendix



\clearpage

\begin{deluxetable}{lcccccc}
\tabletypesize{\small}
\tablecaption{Snapshot Target List\label{survey}}
\tablewidth{0 pt}
\tablehead{
\colhead{Star} & \colhead{Period}  
  & \colhead{$\langle V\rangle$} & \colhead{$E(B-V)$} &
  \colhead{Distance}  & \colhead{Inner Limit\tablenotemark{a}} & 
  \colhead{Outer Limit\tablenotemark{b}} \\
\colhead{} & \colhead{[days]}  
  & \colhead{[mag]} & \colhead{[mag]} &
\colhead{[pc]}  & \colhead{[AU]} & \colhead{[$''$]} 
}
\startdata
  U Aql  &   7.02  &    6.47   &       0.35     &   613 &  184  &  33\\
  TT Aql &   13.75  &   7.14   &      0.49      &   925 &  277  &  22 \\
  FF Aql  &  4.47  &    5.37   &       0.22     &  365 &  109 &   55 \\
 V496 Aql &  6.81  &    7.75    &      0.41     &  989 &   297 &   20 \\
 V1344 Aql &  7.48  &    7.77   &      0.57     &  810 &   243  &  25 \\
 $\eta$ Aql  &  7.18  &    3.90   &    0.12     & 273  &  82 &   73  \\
  RT Aur  &  3.73   &   5.45   &        0.05    & 454&   136 &   44 \\
  CO Aur\tablenotemark{c} & 2.51 & 7.71 & 0.23  & 796 &  239 &   25 \\
  RX Cam  &  7.91  &    7.65   &       0.63     &  715 &  215  &  28 \\
  Y Car  &   3.63  &    8.08   &        0.08    & 1468 &  440 &   14 \\
  $\ell$ Car  &   35.55  &   3.72   &   0.17    &   506 &  152  &  40  \\
  SU Cas\tablenotemark{d} &   2.74  &   5.99    &       023  &    376 &  113 &  53 \\
  TU Cas   &  2.14  &    7.73  &        0.11    &  900 &  270  &  22  \\
 V636 Cas &  8.38  &    7.20   &       0.70     &   535 &  160  &  37 \\
  V Cen  &   5.49  &    6.84   &       0.29     &    709 &  213 &   28 \\
 V553 Cen &  2.06  &    8.46   &       0.22     &    1038 &  311  &  19 \\
 V659 Cen  & 5.62  &    6.67  &        0.21     &   753 &  226 &   27 \\
 V737 Cen  & 7.07   &   6.72  &        0.22     &   848 &  255 &   24 \\
 IR Cep\tablenotemark{d} & 2.98 & 7.78 & 0.43   &  650 & 195 & 31 \\
 $\delta$ Cep &  5.37  &    3.95   &      0.09  &    255 &   76  &  78 \\
 AV Cir\tablenotemark{d} & 4.35  &  7.44 & 0.40 & 701 & 210 & 29 \\
  AX Cir &   5.27   &   6.10  &        0.25     &   527 &  158  &  38 \\
 BP Cir\tablenotemark{d} &  3.39 &  7.71 & 0.32 & 798 & 240 & 25 \\
  R Cru  &   5.83   &   6.77   &       0.19     &   829 &  249 &   24 \\
  S Cru   &   4.69  &    6.60   &       0.16    &  724 &  217  &  28 \\
  T Cru   &  6.73  &    6.57   &       0.19     &   811 &  243 &   25 \\
 BG Cru\tablenotemark{d} & 4.76 & 5.49 &  0.05  & 521 & 156 & 38 \\
  X Cyg  &   16.39 &    6.39   &       0.29     &   981 &  294  &  20 \\
  SU Cyg &   3.85   &   6.90    &      0.08     &   857 &  257  &  23 \\
 DT Cyg\tablenotemark{d} & 3.53 & 5.77 &  0.04  & 521 & 156 & 38 \\
 V1334 Cyg\tablenotemark{d} & 4.74 & 5.98 & 0.07 & 630 & 189 & 32 \\
 $\beta$ Dor &  9.84  &    3.73    &      0.04  &   335 &  100  &  60 \\
  W Gem  &   7.91   &   6.95    &      0.28     &    905 &  272 &   22 \\
 $\zeta$ Gem &  10.15   &   3.92   &      0.02  &   383 &  115 &   52 \\
 V473 Lyr\tablenotemark{c} & 2.62 & 6.18 & 0.03 & 553 & 166 & 36 \\
  T Mon  &   27.02   &   6.14   &       0.14    &  1416 &  425  &  14 \\
  R Mus  &   7.51   &   6.30    &      0.12  	&    844 &  253  &  24 \\
  S Mus  &   9.65  &    6.20    &      0.21  	&   789 &  237  &  25 \\
  S Nor &    9.75   &   6.43   &       0.19  	&   910 &  273  &  22 \\
  Y Oph  &   17.13  &   6.17   &       0.65  	&   510 &  153 &   39 \\
  BF Oph  &  4.07   &   7.34   &       0.25  	&   823 &  247 &   24 \\
  AW Per &   6.46  &    7.55  &       0.53      &     726 &  218  &  28 \\
 V440 Per\tablenotemark{d} & 10.94 & 6.28 & 0.27 & 791 & 237 & 25 \\
 RS Pup & 41.39 & 6.95 & 0.45                   & 1543     & 463 & 13 \\
 AP Pup & 5.08 & 7.37 & 0.21 & 990.             & 297 & 20 \\
 MY Pup\tablenotemark{d} & 8.20 & 5.68 & 0.06   & 728 & 218 & 27 \\
  S Sge  &   8.38  &    5.62   &       0.13     &   641 &  192  &  31 \\
  U Sgr  &   6.75  &    6.70   &       0.40     &   617 &  185  &  32 \\
  W Sgr  &   7.59   &   4.68   &       0.11     &    409 &  123 &   49 \\
  X Sgr  &   7.01   &   4.55   &      0.20      &  321  &  96  &  62 \\
  Y Sgr  &   5.77  &    5.74   &       0.20     &  505 &  152 &   40 \\
  AP Sgr &   5.06   &   6.96   &       0.19     &   845 &  253  &  24 \\
  BB Sgr  &  6.64  &    6.95   &       0.28     &  831 &  249 &   24 \\
 V350 Sgr &  5.15  &    7.52    &      0.32     &  896 &  269  &  22 \\
  RV Sco  &  6.06  &    7.04   &       0.34     &   753 &  226 &   27 \\
 V482 Sco &  4.53  &    7.96   &       0.36     &   968 &  290 &   21 \\
 V636 Sco &  6.79  &    6.65   &       0.20     &  832 &  249  &  24 \\
 V950 Sco\tablenotemark{d} & 4.82 & 7.30 & 0.27 & 849 & 255 & 24 \\
 EW Sct & 5.82 & 7.98 & 1.13                    & 323 & 97 & 62 \\
 SZ Tau\tablenotemark{d} & 4.48 & 6.53 &  0.29  & 557 & 167 & 36 \\
  R TrA  &   3.39   &   6.66   &       0.13  	&  666 &  200 &   30 \\
  S TrA  &   6.32   &   6.40    &      0.10  	&  839 &  252 &   24 \\
  U TrA  &   2.57  &    7.88   &       0.09  	&  1089 &  327 &   18 \\
  LR TrA\tablenotemark{d} & 3.49 & 7.81 & 0.28  &  904 &  271 &   22 \\
  V Vel  &   4.37  &    7.59   &       0.21     &  1018 &  306  &  20 \\
 AH Vel\tablenotemark{d} & 6.04 & 5.70 &  0.07  & 624 & 187 & 32 \\
  T Vul  &   4.43  &    5.76    &      0.06     & 561 &   168  &  36 \\ 
  U Vul   &  7.99  &    7.13   &       0.65     &  548 &   164 &   36 \\ 
  X Vul  &   6.32   &   8.85   &       0.85     &   785 &  236  &  25 \\
  SV Vul  &  44.99  &  7.22    &      0.57      &  1503 &  451  &  13 \\
\enddata
\tablenotetext{a}{Minimum linear separation at which companion could be
detected, corresponding to 0\farcs3 angular separation.}
\tablenotetext{b}{Angular separation corresponding to Galactic tidal limit of
0.1 pc linear separation}
\tablenotetext{c}{Adopted pulsation period is discussed in text}
\tablenotetext{d}{First-overtone pulsator; period has been fundamentalized}
\end{deluxetable}

\begin{deluxetable}{cccccccc}
\tabletypesize{\footnotesize}
\tablecaption{Most Probable Physical Companions\label{prob}}
\tablewidth{0pt}
\tablehead{
 \colhead{$V$} & \colhead{$V-I$} &
  \colhead{Sep.}  & \colhead{PA} & \colhead{Sep.} & \colhead{X-ray} & 
  \colhead{$(V-I)_0$}  & \colhead{No. of} \\
 \colhead{[mag]} & \colhead{[mag]} &
  \colhead{[$''$]}  & \colhead{ [$^\circ$]} & \colhead{[AU]} & 
  \colhead{Detection?}  & \colhead{[mag]}  & \colhead{Companions\tablenotemark{a}}
}
\startdata
\multicolumn{8}{c}{TT Aql} \\
   19.22  & 2.05 &  5.2 & 77  &  4,810 & $\dots$ & 1.49 & 5  \\
   18.77  & 1.92 &  6.6 & 51  & 6,100 & $\dots$ & 1.36 & 5 \\
\multicolumn{8}{c}{FF Aql}\\
11.22  & 0.85 &  6.9 & 146  & 2,520  & $\dots$  & 0.60 & 1 \\
\multicolumn{8}{c}{R Cru}\\
16.281 & 1.17 & 7.64 & 302  & 6,330 & ? & 0.95 & 1 \\
\multicolumn{8}{c}{S Mus}\\
17.94  & 1.56 &  5.0 & 182 & 3,940 & ? & 1.32 & 1 \\
\multicolumn{8}{c}{AP Sgr}\\
17.85  & 1.72 &  6.3 &  86 & 5,320 & $\dots$  & 1.50 & 2 \\
\multicolumn{8}{c}{RV Sco}\\
12.68  & 0.63 &  6.0 &  323  & 4,520 & $\dots$  & 0.24 & 4  \\
\multicolumn{8}{c}{V737 Cen}\\
17.22  & 1.61 &  7.3 & 295 & 6,190 & No & 1.36 & 2 \\
\multicolumn{8}{c}{R Mus}\\
15.68  & 1.17 &  6.9 & 328 & 5,820 & No & 1.03 & 1 \\
\multicolumn{8}{c}{W Sgr}\\
16.10  & 1.75 &  6.3 & 341 & 2,580 & No & 1.62 & 2 \\
\multicolumn{8}{c}{Y Sgr}\\
17.06  & 1.85 & 10.6 & 204 & 5,350 & No & 1.62 & 1 \\
\enddata
\tablenotetext{a}{Total number of candidate companions of Cepheid listed in
Table~A1.}
\end{deluxetable}

\begin{deluxetable}{lcc}
\tabletypesize{\footnotesize}
\tablecaption{Cepheids Nearer than 600 pc\label{near}}
\tablewidth{0pt}
\tablehead{
\colhead{Star} & \colhead{Companion} & \colhead{Companion} \\
\colhead{} & \colhead{{\it HST}/WFC3 $>\!5''$?} & \colhead{{\it XMM/ROSAT\/}?} 
}
\startdata
 FF Aql   & Yes & $\dots$ \\
 $\eta$ Aql & $\dots$ & $\dots$ \\
 RT Aur & $\dots$ & $\dots$\\
 $\ell$ Car    &     Yes &   No \\ 
SU Cas  & $\dots$ & $\dots$ \\
 V636 Cas  &  Yes & $\dots$ \\
 $\delta$ Cep  & see text & $\dots$ \\
AX Cir  & $\dots$ & $\dots$ \\  
 BG Cru  & $\dots$ & $\dots$ \\
DT Cyg  & $\dots$ & $\dots$ \\
 $\beta$ Dor  & $\dots$ & $\dots$ \\
$\zeta$ Gem  & $\dots$ & $\dots$ \\
 V473 Lyr &    Yes &   No  \\
 Y Oph    &     Yes &   No \\
 W Sgr    &     Yes & No \\
 X Sgr  & $\dots$ & $\dots$ \\
 Y Sgr      &   Yes &   No \\
  EW Sct    &  $\dots$  & $\dots$   \\
 SZ Tau  & $\dots$ & $\dots$ \\
 T Vul  &    $\dots$  & $\dots$ \\
 U Vul  & $\dots$ & $\dots$ \\
\enddata
\end{deluxetable}

\clearpage

\section{Appendix A}


The candidate companions of the Cepheids in the survey (\S3.2) are listed 
here.  The columns in the table are the Vega-scale F621M magnitudes and
$F621M-F845M$ colors, followed by these values transformed to ground-based $V$
and $V-I$.  The final three  columns are the separation from the Cepheid  in
arcseconds, the position angle in degrees, and  the separation in AU using the
distance in Table~1.  Table~A1 lists companions which are $\geq$5$\arcsec$ from
the Cepheid;  Table~A2 contains  possible companions $<$5$\arcsec$ from the
Cepheid.  Table~A2 will be discussed  primarily in Paper~III, but is included
here because the search techniques are  the same as those in Table~A1. 

\clearpage

\begin{deluxetable}{ccccccc}
\tabletypesize{\small}
\tablecaption{Candidate Companions of Galactic Cepheids\label{poss}}
\tablenum{A1}
\tablewidth{0pt}
\tablehead{
\colhead{$F621M$} & \colhead{$F621M-F845M$}  
  & \colhead{$V$} & \colhead{$V-I$} &
\colhead{Sep.}  & \colhead{Position} & \colhead{Sep.} \\
\colhead{[mag]} & \colhead{[mag]}  
  & \colhead{[mag]} & \colhead{[mag]} &
\colhead{[$''$]}  & \colhead{Angle [$^\circ$]} & \colhead{[AU]} 
}
\startdata
\multicolumn{7}{c}{TT Aql} \\
  17.83 $\pm$   0.05 & 1.66 $\pm$ 0.05 & 18.49  & 2.07 & 14.1 &       129  & 13,000  \\
  18.13 $\pm$   0.05 & 1.69 $\pm$ 0.05 & 18.79  & 2.09 &  9.6 &    155    & 8,880  \\
  17.74 $\pm$   0.04 & 1.27 $\pm$ 0.04 & 18.41  & 1.73 & 12.5 &     267  & 11,600  \\
  18.56 $\pm$   0.09 & 1.63 $\pm$ 0.09 & 19.22  & 2.05 &  5.2 &       77   &  4,810  \\
  18.11 $\pm$   0.06 & 1.48 $\pm$ 0.06 & 18.77  & 1.92 &  6.6 &       51    & 6,100  \\
\multicolumn{7}{c}{FF Aql} \\
  10.81 $\pm$   0.00 & 0.55 $\pm$ 0.00 & 11.22  & 0.85 &  6.9 &     146  & 2,520  \\
\multicolumn{7}{c}{V496 Aql} \\
  16.57 $\pm$   0.02 & 1.18 $\pm$ 0.02 & 17.25  & 1.64 &  6.7 &     85  & 6,620  \\
  14.81 $\pm$   0.01 & 0.84 $\pm$ 0.01 & 15.41  & 1.26 & 20.9 &       260   & 20,600 \\
\multicolumn{7}{c}{Y Car} \\
  18.02 $\pm$   0.05 & 1.28 $\pm$ 0.05 & 18.70  & 1.74 & 11.6 &     6  & 17,000  \\
\multicolumn{7}{c}{$\ell$ Car}\\
  14.67 $\pm$   0.01 & 0.71 $\pm$ 0.01 & 15.21  & 1.09 & 19.1 &   10   & 9,660  \\
\multicolumn{7}{c}{V636 Cas}\\
  16.08 $\pm$   0.02 & 1.29 $\pm$ 0.02 & 16.76  & 1.75 & 19.9 &   153	& 10,600  \\
\multicolumn{7}{c}{V Cen}\\
  14.97 $\pm$   0.01 & 0.87 $\pm$ 0.01 & 15.59  & 1.30 & 13.4 &   102  & 9,500   \\
\multicolumn{7}{c}{V659 Cen}\\
  15.29 $\pm$   0.01 & 0.79 $\pm$ 0.01 & 15.87  & 1.20 & 22.1 &   335	& 16,600   \\
  17.42 $\pm$   0.04 & 1.12 $\pm$ 0.04 & 18.09  & 1.59 & 14.7 &  341  & 11,100   \\
  15.97 $\pm$   0.02 & 0.84 $\pm$ 0.02 & 16.57  & 1.27 & 20.2 &  238  & 15,200  \\
  15.16 $\pm$   0.01 & 0.77 $\pm$ 0.01 & 15.73  & 1.17 & 23.8 &  81 & 17,900  \\
  16.93 $\pm$   0.03 & 1.01 $\pm$ 0.03 & 17.59  & 1.47 & 17.0 &  59 & 12,800  \\
\multicolumn{7}{c}{V737 Cen}\\
  16.55 $\pm$   0.03 & 1.14 $\pm$ 0.03 & 17.22  & 1.61 &  7.3 &  295 & 6,190   \\
  17.00 $\pm$   0.03 & 1.14 $\pm$ 0.03 & 17.67  & 1.61 & 17.1 &       231 & 14,500  \\
\multicolumn{7}{c}{IR Cep}\\
  16.95  $\pm$  0.03 & 1.36 $\pm$ 0.04 & 17.62  & 1.81 & 18.1 &  42 & 11,800   \\ 
\multicolumn{7}{c}{AV Cir}\\
   13.18  $\pm$ 0.01 & 0.71 $\pm$ 0.01 & 13.72  & 1.10 & 17.4 & 133  & 12,200	    \\ 
\multicolumn{7}{c}{BP Cir}\\
  16.48 $\pm$   0.02 & 0.98 $\pm$ 0.02 & 17.13  & 1.43 & 14.1 & 296  & 11,200	\\
  16.92 $\pm$   0.03 & 1.23 $\pm$ 0.03 & 17.60  & 1.70 & 21.2 &  271  & 16,900  \\
  17.77 $\pm$   0.04 & 1.39 $\pm$ 0.04 & 18.44  & 1.84 & 11.6 &  168 & 9,260  \\
\multicolumn{7}{c}{R Cru}\\
   15.71 $\pm$  0.02 & 0.77 $\pm$ 0.03 & 16.281 & 1.17 & 7.64 &  302  & 6,330	\\
\multicolumn{7}{c}{S Cru}\\
  15.91 $\pm$   0.02 & 1.11 $\pm$ 0.02 & 16.59  & 1.58 & 13.8 &  70  & 9,990   \\
  17.23 $\pm$   0.04 & 1.09 $\pm$ 0.04 & 17.90  & 1.55 & 11.9 &       20 & 8,620  \\
\multicolumn{7}{c}{T Cru}\\
  17.39 $\pm$   0.04 & 1.21 $\pm$ 0.04 & 18.07  & 1.67 &  9.1 &  134 & 7,380   \\
\multicolumn{7}{c}{X Cyg}\\
  17.94 $\pm$   0.06 & 1.21 $\pm$ 0.06 & 18.61  & 1.68 & 12.9 &       96  & 12,700   \\
  15.70 $\pm$   0.02 & 0.91 $\pm$ 0.02 & 16.33  & 1.36 & 14.8 &     298 & 14,500   \\
\multicolumn{7}{c}{SU Cyg}\\
  16.30 $\pm$   0.02 & 0.81 $\pm$ 0.02 & 16.90  & 1.23 & 25.2 &  87  & 21,600	\\
\multicolumn{7}{c}{W Gem}\\
  16.24 $\pm$   0.02 & 0.82 $\pm$ 0.02 & 16.84  & 1.24 & 14.4 &       58 & 13,000   \\
\multicolumn{7}{c}{V473 Lyr}\\
  14.30 $\pm$   0.01 & 0.81 $\pm$ 0.01 & 14.89  & 1.23 & 15.0 & 44  & 8,300   \\
\multicolumn{7}{c}{T Mon}\\
  17.14 $\pm$   0.05 & 0.80 $\pm$ 0.05 & 17.73  & 1.21 &  6.5 & 276  & 9,200   \\
  16.14 $\pm$   0.02 & 0.58 $\pm$ 0.02 & 16.59  & 0.91 & 18.5 & 166 & 26,200  \\
\multicolumn{7}{c}{R Mus}\\
  15.11 $\pm$   0.01 & 0.77 $\pm$ 0.01 & 15.68  & 1.17 &  6.9 & 328  & 5,820   \\
\multicolumn{7}{c}{S Mus}\\
  17.27 $\pm$   0.05 & 1.09 $\pm$ 0.05 & 17.94  & 1.56 &  5.0 & 182  & 3,940   \\
\multicolumn{7}{c}{S Nor} \\
  13.51 $\pm$   0.01 & 0.58 $\pm$ 0.01 & 13.95  & 0.90 & 14.6 &   289  & 13,300  \\
  15.89 $\pm$   0.02 & 0.76 $\pm$ 0.02 & 16.45  & 1.15 & 19.8 &  172   & 18,000  \\
  15.73 $\pm$   0.02 & 0.79 $\pm$ 0.02 & 16.32  & 1.20 & 20.1 &   188 & 18,300  \\
  17.41 $\pm$   0.04 & 0.98 $\pm$ 0.04 & 18.06  & 1.44 &  8.5 &   44 & 7,740  \\
  16.70 $\pm$   0.03 & 1.08 $\pm$ 0.03 & 17.37  & 1.54 & 13.5 &  7 & 12,300  \\
  17.33 $\pm$   0.04 & 1.06 $\pm$ 0.04 & 18.00  & 1.52 & 15.0 &   1 & 13,600  \\
\multicolumn{7}{c}{Y Oph}\\
  16.47 $\pm$   0.03 & 1.57 $\pm$ 0.03 & 17.13  & 2.00 & 18.1 &       212  & 9,230   \\
\multicolumn{7}{c}{BF Oph}\\
  17.04 $\pm$   0.03 & 1.12 $\pm$ 0.03 & 17.71  & 1.58 & 18.8 & 244   & 15,500   \\
\multicolumn{7}{c}{V440 Per}\\
  15.13 $\pm$   0.01 & 0.80 $\pm$ 0.01 & 15.72  & 1.21 & 10.9 &  305 & 8,620   \\
  13.33 $\pm$   0.00 & 0.65 $\pm$ 0.00 & 13.83  & 1.01 & 10.6 & 131  & 8,380  \\
\multicolumn{7}{c}{RS Pup}\\
  18.13 $\pm$   0.06 & 1.16 $\pm$ 0.06 & 18.81  & 1.63 & 10.8 &       317  & 16,700   \\
  16.21 $\pm$   0.02 & 0.85 $\pm$ 0.02 & 16.82  & 1.28 & 21.8 & 54  & 33,600  \\
\multicolumn{7}{c}{U Sgr}\\
  15.64 $\pm$   0.02 & 1.06 $\pm$ 0.02 & 16.31  & 1.52 & 19.4 &       18 & 12,000   \\
  16.76 $\pm$   0.03 & 1.52 $\pm$ 0.03 & 17.42  & 1.95 & 13.9 &  127  & 8,580  \\
  17.00 $\pm$   0.03 & 1.51 $\pm$ 0.03 & 17.66  & 1.94 & 17.1 &       164  & 10,500  \\
\multicolumn{7}{c}{W Sgr}\\
  15.43 $\pm$   0.03 & 1.28 $\pm$ 0.03 & 16.10  & 1.75 &  6.3 &  341  & 2,580	\\
  15.76 $\pm$   0.03 & 1.31 $\pm$ 0.03 & 16.44  & 1.77 & 19.0 &       46 & 7,770  \\
\multicolumn{7}{c}{Y Sgr} \\
  16.39 $\pm$   0.03 & 1.40 $\pm$ 0.03 & 17.06  & 1.85 & 10.6 &       204  & 5,350   \\
\multicolumn{7}{c}{AP Sgr}\\
  17.47 $\pm$   0.04 & 1.16 $\pm$ 0.04 & 18.15  & 1.63 & 23.5 &  258 & 19,900	\\
  17.18 $\pm$   0.04 & 1.25 $\pm$ 0.04 & 17.85  & 1.72 &  6.3 &       86 & 5,320  \\
\multicolumn{7}{c}{V350 Sgr}\\
  15.42 $\pm$   0.01 & 1.11 $\pm$ 0.02 & 16.09  & 1.58 & 10.6 &     156. & 9,500   \\
  16.26 $\pm$   0.02 & 1.03 $\pm$ 0.02 & 16.92  & 1.49 & 18.0 &       330  & 16,100  \\
  18.06 $\pm$   0.05 & 1.37 $\pm$ 0.05 & 18.73  & 1.83 & 11.4 &  270 & 10,200  \\
  15.41 $\pm$   0.01 & 0.81 $\pm$ 0.01 & 16.00  & 1.22 & 10.6 &  66  & 9,500  \\
\multicolumn{7}{c}{RV Sco}\\
  12.37 $\pm$   0.00 & 0.41 $\pm$ 0.00 & 12.68  & 0.63 &  6.0 &  323  & 4,520  \\
  17.52 $\pm$   0.04 & 1.24 $\pm$ 0.04 & 18.19  & 1.71 & 14.8 &  79  & 11,100 \\
  17.12 $\pm$   0.03 & 1.09 $\pm$ 0.03 & 17.79  & 1.56 & 10.6 &  187  & 7,980  \\
  16.66 $\pm$   0.03 & 1.33 $\pm$ 0.03 & 17.33  & 1.79 & 23.0 &  192  & 17,300 \\
\multicolumn{7}{c}{V636 Sco}\\
  15.91 $\pm$   0.02 & 0.79 $\pm$ 0.02 & 16.49  & 1.20 & 14.4 &  268  & 12,000  \\
  15.26 $\pm$   0.01 & 0.76 $\pm$ 0.01 & 15.83  & 1.17 & 18.0 &  215  & 15,000 \\
\multicolumn{7}{c}{V950 Sco}\\
  17.64 $\pm$   0.04 & 1.34 $\pm$ 0.04 & 18.31  & 1.80 & 19.9 &   180  & 16,800  \\
  17.32 $\pm$   0.04 & 1.24 $\pm$ 0.04 & 17.99  & 1.71 & 11.9 &   331	& 10,100 \\
  15.20 $\pm$   0.01 & 0.72 $\pm$ 0.01 & 15.74  & 1.11 & 15.9 &   316  & 13,500 \\
  16.94 $\pm$   0.03 & 1.01 $\pm$ 0.03 & 17.60  & 1.47 & 14.3 &    312     & 12,100 \\
  16.22 $\pm$   0.02 & 0.86 $\pm$ 0.02 & 16.84  & 1.29 & 19.6 &   280  & 16,600 \\
  17.21 $\pm$   0.03 & 1.07 $\pm$ 0.03 & 17.88  & 1.53 & 24.6 &   116  & 20,800 \\
\multicolumn{7}{c}{S TrA}\\
  15.42 $\pm$   0.01 & 0.70 $\pm$ 0.01 & 15.95  & 1.07 & 17.6 &   0   & 14,800  \\
  15.44 $\pm$   0.01 & 0.91 $\pm$ 0.01 & 16.08  & 1.35 & 15.0 &  48	& 12,600  \\
\multicolumn{7}{c}{U TrA}\\
  17.16 $\pm$   0.03 & 0.79 $\pm$ 0.03 & 17.74  & 1.20 & 20.5 &  360  & 22,300  \\
  16.70 $\pm$   0.03 & 0.85 $\pm$ 0.03 & 17.31  & 1.28 & 13.4 &  256  & 14,600 \\
\multicolumn{7}{c}{V Vel}\\
  13.77 $\pm$   0.01 & 0.40 $\pm$ 0.01 & 14.08  & 0.63 & 16.8 &  65  & 17,100  \\
  17.84 $\pm$   0.04 & 0.99 $\pm$ 0.04 & 18.49  & 1.44 & 16.5 &  56  & 16,800 \\
\multicolumn{7}{c}{SV Vul}\\
  18.37 $\pm$   0.06 & 1.35 $\pm$ 0.06 & 19.04  & 1.80 & 15.1 &   241  & 22,700  \\
  18.17 $\pm$   0.05 & 1.12 $\pm$ 0.05 & 18.84  & 1.59 & 16.9 &  131  & 25,400 \\
  18.55 $\pm$   0.07 & 1.20 $\pm$ 0.07 & 19.23  & 1.67 & 20.9 &  68  & 31,400 \\
  14.50 $\pm$   0.01 & 0.81 $\pm$ 0.01 & 15.09  & 1.22 & 15.1 &  86   & 22,700 \\
\enddata
\end{deluxetable}

\clearpage

\begin{deluxetable}{ccccccc}
\tablecaption{Candidate Companions with Separation $<$5$\arcsec$\label{poss}}
\tablenum{A2}
\tablewidth{0pt}
\tablehead{
\colhead{$F621M$} & \colhead{$F621M-F845M$}  
  & \colhead{$V$} & \colhead{$V-I$} &
\colhead{Sep.}  & \colhead{Position} & \colhead{Sep.} \\
\colhead{[mag]} & \colhead{[mag]}  
  & \colhead{[mag]} & \colhead{[mag]} &
\colhead{[$''$]}  & \colhead{Angle [$^\circ$]} & \colhead{[AU]} 
}
\startdata
\multicolumn{7}{c}{TT Aql}\\
  17.16 $\pm$   0.05   & 1.40 $\pm$ 0.05 & 17.82  & 1.85 & 3.8 &    67   & 3,520  \\
\multicolumn{7}{c}{V496 Aql}\\
  18.39 $\pm$   0.09   & 1.56 $\pm$ 0.09 & 19.05  & 1.99 & 4.3 &    8    & 4,250  \\
\multicolumn{7}{c}{Y Car}\\
  16.45 $\pm$   0.02   & 0.60 $\pm$ 0.02 & 16.91  & 0.93 & 2.6 &   55  & 3,820 \\
  16.79 $\pm$   0.03   & 0.86 $\pm$ 0.03 & 17.40  & 1.29 & 3.2 &  112	& 4,700 \\
\multicolumn{7}{c}{BB Sgr}\\
   17.41 $\pm$  0.06   & 1.31 $\pm$ 0.07 & 18.081 & 1.77 & 3.3 &  158  & 2,740	\\
\multicolumn{7}{c}{V350 Sgr}\\
  17.24 $\pm$   0.04   & 1.31 $\pm$ 0.04 & 17.91  & 1.77 & 3.1 &  128 & 2,780  \\
\multicolumn{7}{c}{RV Sco}\\
  15.37 $\pm$   0.01   & 0.90 $\pm$ 0.01 & 16.00  & 1.34 & 3.6 &  173  & 2,710 \\
\enddata
\end{deluxetable}

\end{document}